\documentclass[twocolumn]{aastex61}

\usepackage{color}
\usepackage{amsmath}

\newcommand\aastex{AAS\TeX}

\def\ba{\begin{eqnarray}}
\def\ea{\end{eqnarray}}

\def\msun{M_\odot}
\def\msol{M_\odot}

\def\te{T_{\rm eff}}

\def\Hp{H_{\rm{p}}}
\def\dov{d_{\mathsf{ov}}}
\def\lov{l_{\mathsf{ov}}}
\def\dpratt{D_{\mathsf{EX}}}

\def\ltsima{$\; \buildrel < \over \sim \;$}
\def\simlt{\lower.5ex\hbox{\ltsima}}
\def\gtsima{$\; \buildrel > \over \sim \;$}
\def\simgt{\lower.5ex\hbox{\gtsima}}

\newcommand{\be}{\begin{equation}}
\newcommand{\eeq}{\end{equation}}

\newenvironment{packed_item}{
\begin{itemize}
 \setlength{\itemsep}{0.5pt}
  \setlength{\parskip}{-1.5pt}
  \setlength{\parsep}{-1pt}
}{\end{itemize}}

\def\ba{\begin{eqnarray}}
\def\ea{\end{eqnarray}}

\received{}
\revised{}
\accepted{}
\submitjournal{ApJ}

\shorttitle{\aastex\ Lithium depletion in solar-like stars}
\shortauthors{Baraffe et al.}
\begin{document}

\title{Lithium depletion in solar-like stars: effect of overshooting based on realistic multi-dimensional simulations}

\correspondingauthor{Isabelle Baraffe}
\email{i.baraffe@ex.ac.uk}

\author{I. Baraffe}
\affiliation{Astrophysics Group, University of Exeter, EX4 4QL Exeter, UK}
\affiliation{ Ecole Normale Sup\'erieure de Lyon, CRAL, UMR CNRS 5574, 69364 Lyon
  Cedex 07, France} 
\author{J. Pratt} 
\affiliation{Astrophysics Group, University of Exeter, EX4 4QL Exeter, UK}
\author{T. Goffrey}
\affiliation{Astrophysics Group, University of Exeter, EX4 4QL Exeter, UK}
\author{T. Constantino }
\affiliation{Astrophysics Group, University of Exeter, EX4 4QL Exeter, UK}
\author{D. Folini }
\affiliation{ Ecole Normale Sup\'erieure de Lyon, CRAL, UMR CNRS 5574, 69364 Lyon
  Cedex 07, France}
\author{M. V. Popov}
\affiliation{ Ecole Normale Sup\'erieure de Lyon, CRAL, UMR CNRS 5574, 69364 Lyon
  Cedex 07, France}
\author{R. Walder}
         \affiliation{ Ecole Normale Sup\'erieure de Lyon, CRAL, UMR CNRS 5574, 69364 Lyon
  Cedex 07, France} 
  \author{M. Viallet} 
\affiliation{Astrophysics Group, University of Exeter, EX4 4QL Exeter, UK}

\begin{abstract}
We study lithium depletion in low-mass and solar-like stars  as a function of time, using a new diffusion coefficient describing extra-mixing taking place at the bottom of a convective envelope. This new form is motivated by multi-dimensional fully compressible, time implicit hydrodynamic simulations performed with the MUSIC code.  Intermittent convective mixing at the convective boundary in a star can be modeled using extreme value theory, a statistical analysis frequently used for finance, meteorology, and environmental science.  
In this letter, we implement this statistical diffusion coefficient in a one-dimensional stellar evolution code, using parameters calibrated from multi-dimensional hydrodynamic simulations of a young low-mass star.
We propose a new scenario that can explain observations of the surface abundance of lithium in the Sun and in clusters covering a wide range of ages, from $\sim$ 50 Myr to $\sim$ 4 Gyr. Because it relies on our physical model of convective penetration, this scenario has a limited number of assumptions.
It can explain the observed trend between rotation and depletion, based on a single additional assumption, namely that rotation affects  the mixing efficiency at the convective boundary. We suggest the existence of a threshold in stellar rotation rate above which rotation strongly prevents the vertical penetration of plumes and below which rotation has small effects. In addition to providing a possible explanation for the long standing problem of lithium depletion in pre-main sequence and main sequence stars, the strength of our scenario is that its basic assumptions can be tested by future hydrodynamic simulations.
\end{abstract}

\keywords{Stars: evolution --- Stars: rotation  --- Stars: solar-type  --- Stars: pre-main sequence ---
             Convection --- Hydrodynamics }

\section{Introduction} \label{intro}
Lithium abundance observed at the surface of a star is a sensitive probe of its interior. Despite decades of theoretical efforts, the observed surface abundance of lithium
in pre-main (PMS) and main sequence (MS) stars in clusters of different ages remains difficult to interpret.  To reproduce the observed mass and age dependence of lithium depletion, models require an increasingly complex combination of mixing processes from overshooting, rotational mixing, and microscopic diffusion.  Several free parameters define these models, which must be adjusted based on observations \citep[see e.g][]{Castro16}.  
Since the nineties, observational evidence accumulates  showing that rotation plays a key role in limiting the lithium depletion process \citep[see e.g][]{Tschaepe01}.
 Fast rotators appear to be less depleted than their slow rotating counterparts and this trend persists
 in clusters of different ages. This is observed at $\sim$ 5 Myr in NGC2264 \citep{Bouvier16}, $\sim$ 25 Myr in the $\beta$ Pictoris association  \citep{Messina16} or at $\sim$ 120 Myr in the Pleiades \citep{Barrado16}. These observations totally counter current models of rotational mixing.
 To explain this trend, a complex link between lithium depletion and rotation needs to be invoked. \cite{Bouvier08} suggest that slow rotators develop a higher level of differential rotation, and thus of Li destruction, while fast rotators exhibit little core-envelope decoupling. This idea, however, still needs to be proven. \citet{Somers14, Somers15} suggest that fast rotation and strong magnetic fields result in radius inflation that can inhibit Li depletion. Though attractive, this idea adds another layer of complexity and additional free parameters in 1D stellar evolution models. Lithium depletion in PMS and MS stars thus remains an open problem.
 
 In this letter, we propose a scenario based on a single physical process to explain the main observed trends of lithium depletion, namely mixing due to the penetration of convective motions in the stable region at the base of the convective envelope of PMS  and  MS stars.  Mixing at convective boundaries, also referred to as overshooting or penetration, is one of the oldest unsolved problems of stellar structure and evolution theory
 \citep[e.g][]{Shaviv73, Schmitt84}.
 Overshoot at the base of the solar convection zone is  crucial to interpreting helioseismology data
\citep{Christensen-Dalsgaard11} and understanding the Sun's magnetic activity \citep{Rempel04}. Many stellar evolution models include convective boundary mixing. A common treatment is to arbitrarily fix an overshoot length $\lov$ which characterises the width of the overshooting layer where mixing is very efficient. The width $\lov$ is calibrated from observations. We recently derived a new form for the diffusion coefficient describing mixing below the convective envelope of a PMS star based on multi-dimensional fully compressible hydrodynamic simulations \citep{Pratt17}. This original approach is based on the statistical analysis of  numerical data using extreme value theory, a well known statistical method in finance, meteorology or environmental science. In this work, we implement this diffusion coefficient in our one-dimensional (1D) stellar evolution code \citep{Baraffe98} and explore its effect on lithium depletion in the stellar mass range 0.85-1.5 $\msol$ for which observations  in various clusters of different ages are available. We demonstrate that the general trend of lithium depletion as a function of age and mass is reproduced, if  one additional assumption is made regarding the effect of rotation on the efficiency of the convective boundary mixing. 
We suggest the existence of a threshold in stellar rotation rate above which rotation strongly prevents the vertical penetration of plumes and below which rotation has small effects. We conclude this preliminary work with future plans to explore this scenario with multi-dimensional stellar simulations.
 
\section{Multi-dimensional simulations of convective overshoot \label{sect2}} 

We have recently explored hydrodynamic simulations of two-dimensional compressible  convection in a pre-main sequence 1 $\msol$ star using a new time-implicit code, the MUltidimensional Stellar Implicit Code (MUSIC) \citep{Viallet11, Viallet13, Viallet16, Geroux16, Goffrey17}. The 2D models are in spherical geometry and are calculated using realistic stellar interior conditions \citep[including a realistic stellar equation of state and opacities, see][]{Pratt16, Pratt17}. These simulations cover up to 525 convective turnovers, producing statistically robust data that characterises the extent and impact of convective penetration at the bottom of the convective envelope. The results reveal the frequency and physical importance of extreme events where intermittent convective plumes penetrate deep below from the convective boundary into the stable radiative region. Note that preliminary 3D simulations show the same patterns as  found in 2D and confirm the existence of extreme plume events.
Associating mixing process with plume penetration, we apply a statistical method based on extreme value theory to derive the cumulative distribution function (CDF) of the maximal penetration depth obtained in our numerical simulations \citep[see][]{Pratt17}. 
We derive a new form for a diffusion coefficient $\dpratt$ characterised by this CDF and describing  mixing driven by the convective plumes in the penetration layer. This formalism is reasonable for stellar interiors characterised by large P\'eclet numbers. 
The diffusion coefficient in the penetration layer is expressed as a function of the radial position $r$ and has the form:

\begin{eqnarray}
\dpratt(r) = D_0 \bigg \{1- \exp \bigg [ -\exp\left(- { { (r_{\rm B} - r) \over R} - \mu \over \lambda} \right) \bigg ] \bigg \}.  
\label{eq1}
\end{eqnarray} 
The parameter $D_0$ represents the mixing in the convective zone and is assumed to be equal to the diffusion coefficient at the bottom of the convective envelope $D_{\mathsf{MLT}} = 1/3 \, L_{\rm mix} \, v_{\mathsf{MLT}}$ defined by mixing length theory (MLT) and calculated from the 1D stellar structure model (see \S \ref{sect3}). In this expression $L_{\rm mix}$ is proportional to the pressure scale-height, $\Hp$, and $v_{\mathsf{MLT}}$ is the convective velocity. $R$ is the total radius of the star and $r_{\rm B}$ is the position of the Schwarzschild boundary at the bottom of the convective envelope. The coefficients $\lambda$ and $\mu$ are obtained from the fit to the cumulative distribution function derived from our numerical simulations \citep[see][]{Pratt17}. In the following we adopt the parameters derived from simulation YS0 of a pre-MS 1 $\msol$ model \citep[see Table 2 of][]{Pratt17}, $\lambda=6 \cdot 10^{-3}$  and $\mu=5 \cdot 10^{-3}$. Simulation YS0 is the most relevant in terms of collection of statistical data as it covers 525 convective turnovers.  
In our exploration of lithium depletion, these parameters and the resulting diffusion coefficient do not depend on the stellar structure or age.  We make this choice for the present study, because hydrodynamic convection data is not yet available throughout the course of the star's evolution.  Extending the range of hydrodynamic simulations is planned in the future (see \S \ref{conclusion}).
There is therefore no model presently available for how mixing at a convective boundary changes in time, as the radial extent of the star's convection zone shrinks.  

\section{Application to 1D stellar evolution models} 
\label{sect3}
\begin{figure}[h]
\vspace{-2cm}
\includegraphics[height=13cm]{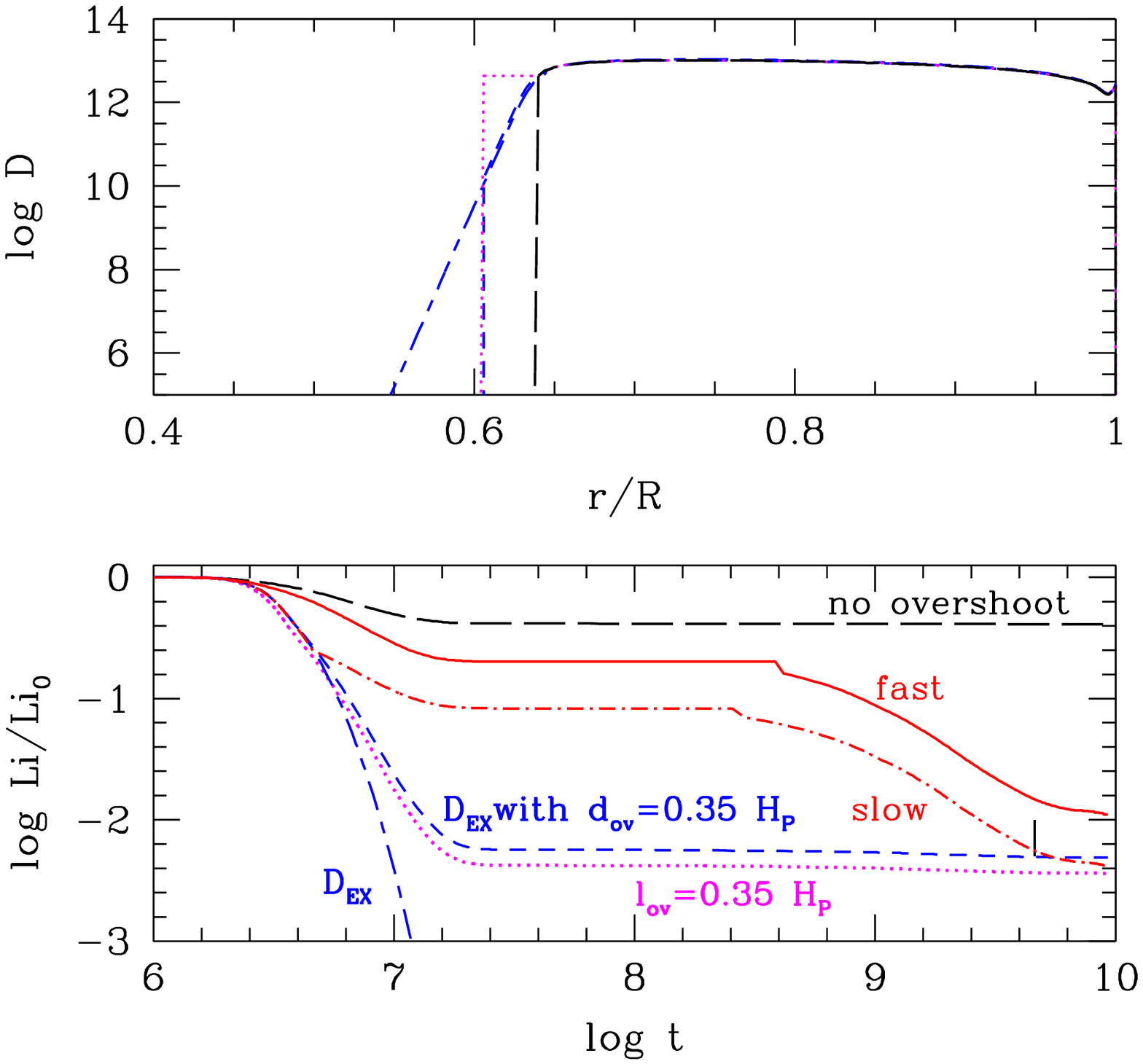}
\vspace{-3cm}
\caption{Radial profile of the diffusion coefficient $D$ (upper panel) and lithium depletion as a function of age (lower panel) in a 1 $\msol$ stellar model with different treatments of overshooting.
{\bf Upper panel}:  radial profile of $D$ (in $\mathrm{cm}^2~\mathrm{s}^{-1}$) as a function of radius $r$ (divided by the total stellar radius $R$) at t=20 Myr. 
Standard model without overshooting: long-dash black;  $\dpratt$ with no limitation
on $\dov$: long-dash - short-dash blue; $\dpratt$ with $\dov=0.35 \Hp$: dash blue; 
simple overshooting length $\ell_{\mathsf{ov}} = 0.35 \Hp$: dot magenta (see \S \ref{sect3}). 
{\bf Lower panel}: abundance of Li is normalised to the initial abundance Li$_0$ and time is in yr. The black and blue curves correspond to  models with the diffusion coefficients displayed in the upper panel (same color and linestyle). Curves in red
are models with assumptions for rotation effects (see details and explanations in \S \ref{sect4}):  fast initial rotation (solid red) and slow initial rotation (dash-dot red). The black vertical line at the Sun's age shows the range  1/100 $\simlt \mathrm{Li/Li}_{0} \simlt$ 1/200 for the solar surface abundance of lithium \citep{Grevesse98}.}

\label{fig_tli}
\end{figure}

We  implement the diffusion coefficient $\dpratt$ from eq.~\eqref{eq1} in our 1D stellar evolution code using the same input physics as in the models of \citet{Baraffe98}. These models are widely used for comparisons with observations of low mass-stars in clusters.  We use a mixing length parameter $L_{\rm mix} = 1.9 \Hp$ to reproduce the Sun's luminosity and radius at 4.6 Gyr.
Figure \ref{fig_tli} shows the evolution of the surface abundance of lithium  for a 1 $\msol$ model with different treatments of mixing in the overshooting layer below the convective envelope\footnote{We adopt  the meteoritic abundance A(Li)=3.3 \citep{Grevesse98} as the initial abundance of Li.}. When extra-mixing is based on the diffusion coefficient $\dpratt$, the depletion of lithium is  too rapid  and it is completely destroyed after around 30 Myr (long-dash - short-dash blue curve).  Such strong depletion is prevented if the layer where mixing proceeds according to $\dpratt$  is limited;  we therefore explore a fixed limit for the overshooting layer using a width $\dov$.  The observed solar lithium abundance is reproduced within the error bars if $\dov \sim 0.30\Hp - 0.35 \Hp$.  In contrast, if a simple overshooting length $\ell_{\mathsf{ov}}$ with efficient mixing\footnote{Assuming $D$ is constant and equal to the value of $D_{\rm MLT}$ at the bottom of the convective envelope.} is adopted, and set to the {\it same value} as this limiting width $\dov$, our stellar evolution calculations show that lithium is more depleted  than the observed value (see dotted curve in Figure \ref{fig_tli}). 
Obviously, one can reproduce the observed Li abundance with a smaller value of $\ell_{\mathsf{ov}}$, but this comparison illustrates the effect of the decaying form of $\dpratt$ (see upper panel of Figure \ref{fig_tli}) which produces a smoother transition between the convective region and the stable one, in better agreement with the smooth overshoot profiles inferred from helioseismology \citep{Christensen-Dalsgaard11, Thevenin17}. 

\begin{figure*}
\gridline{\fig{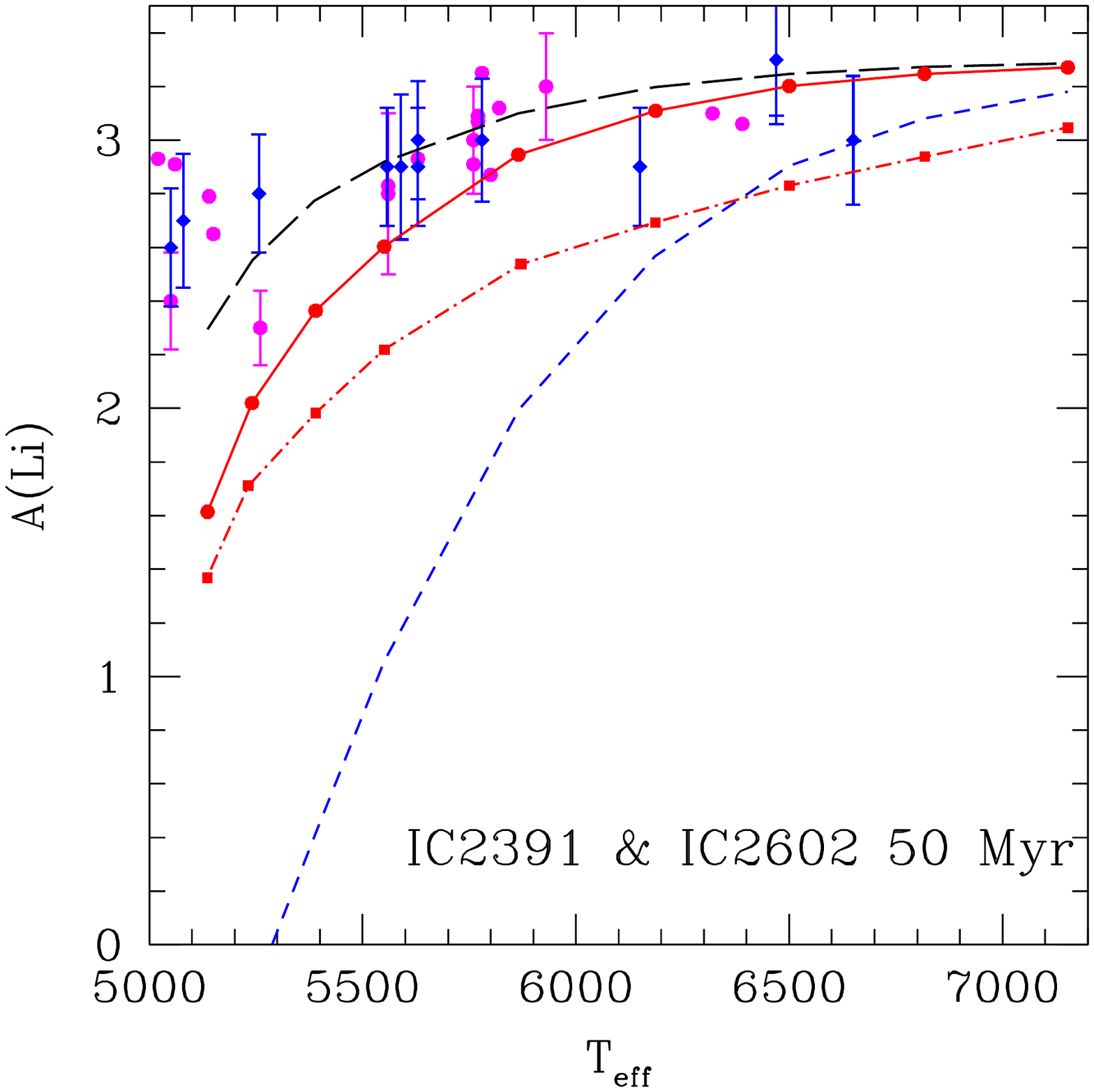}{0.35\textwidth}{\vspace{-1.5cm} (a)}
          \fig{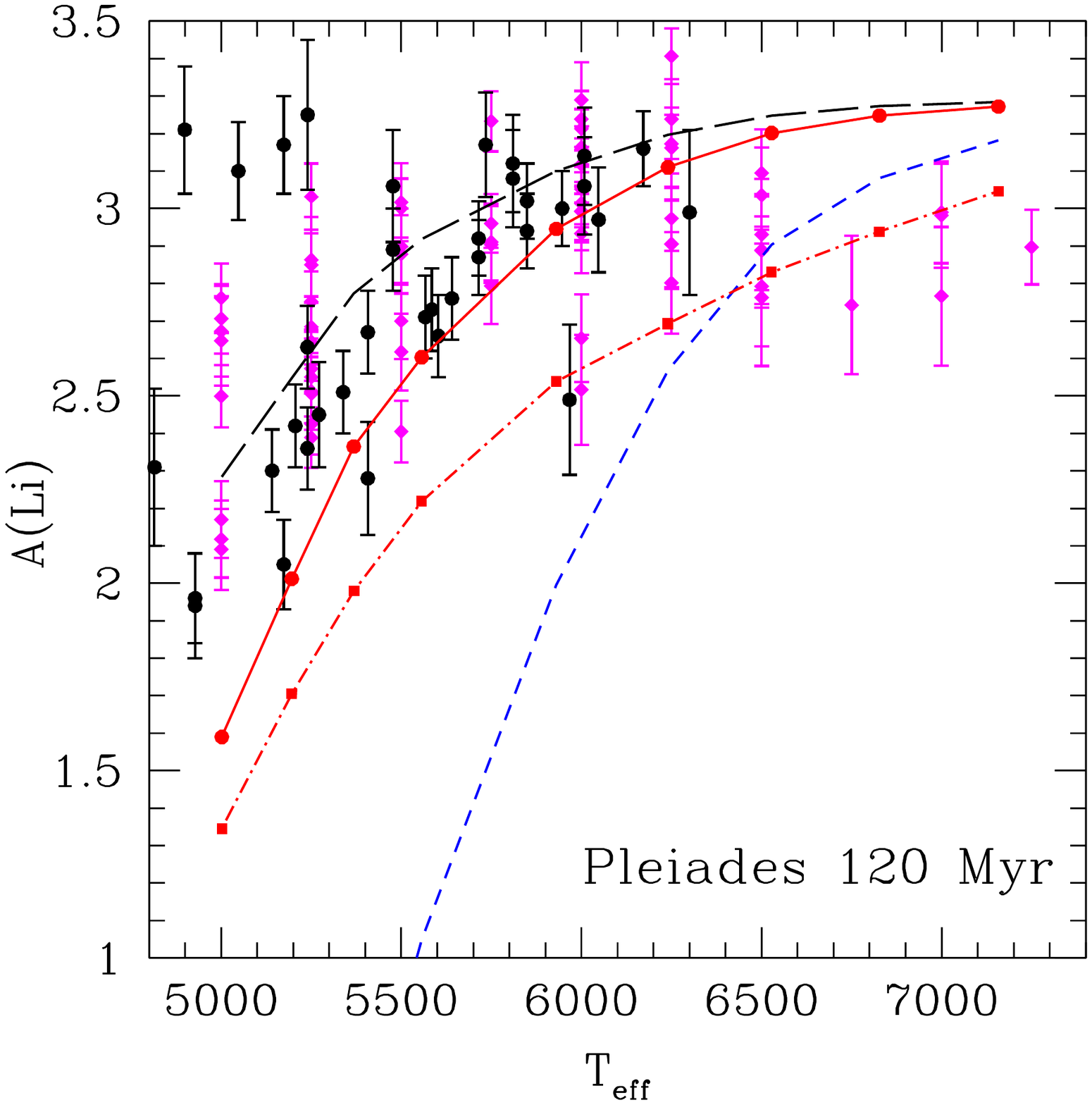}{0.35\textwidth}{\vspace{-1.5cm} (b)}
          \fig{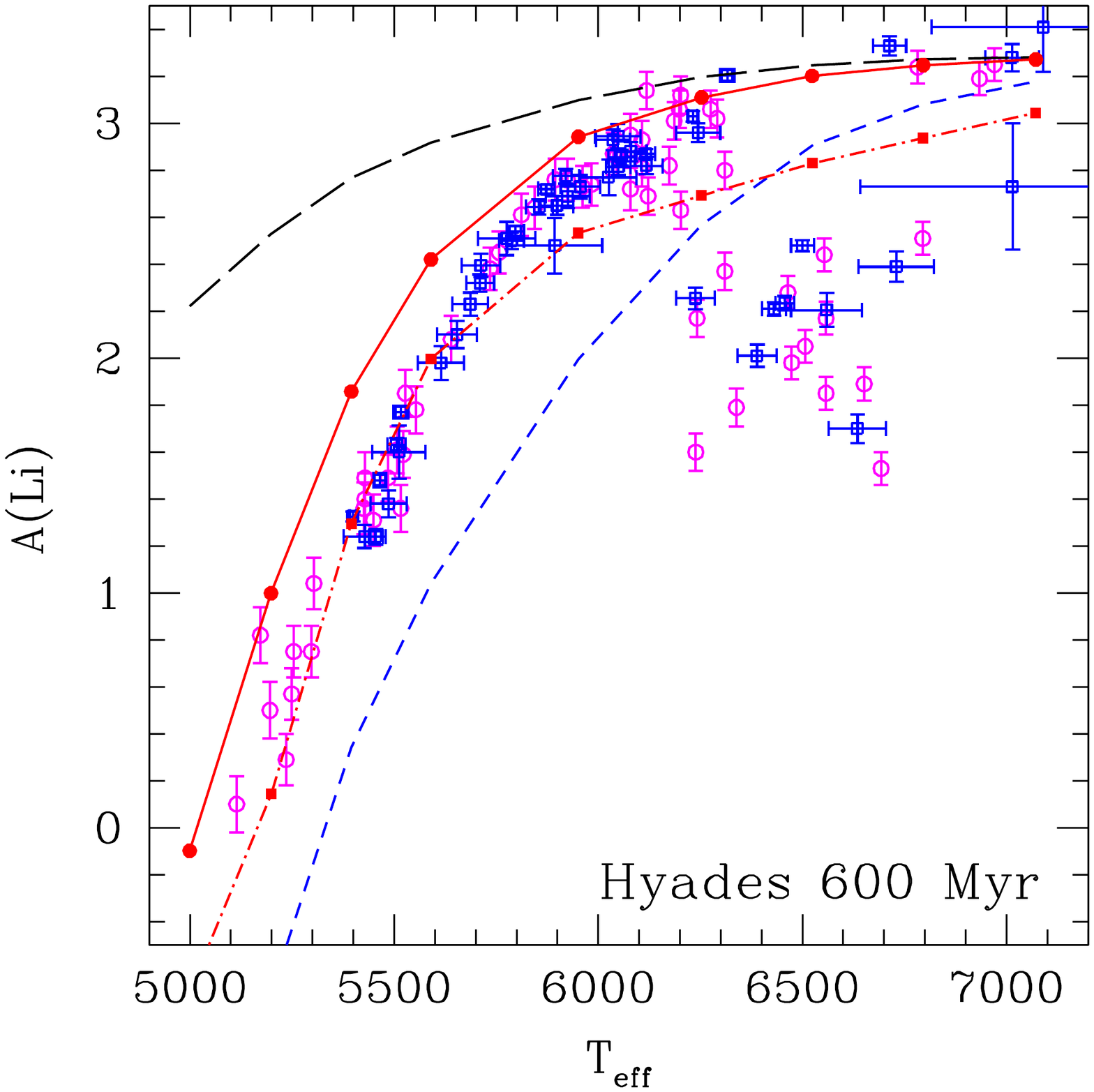}{0.35\textwidth}{\vspace{-1.5cm} (c)}}
          \vspace{-2.5cm}
\gridline{\fig{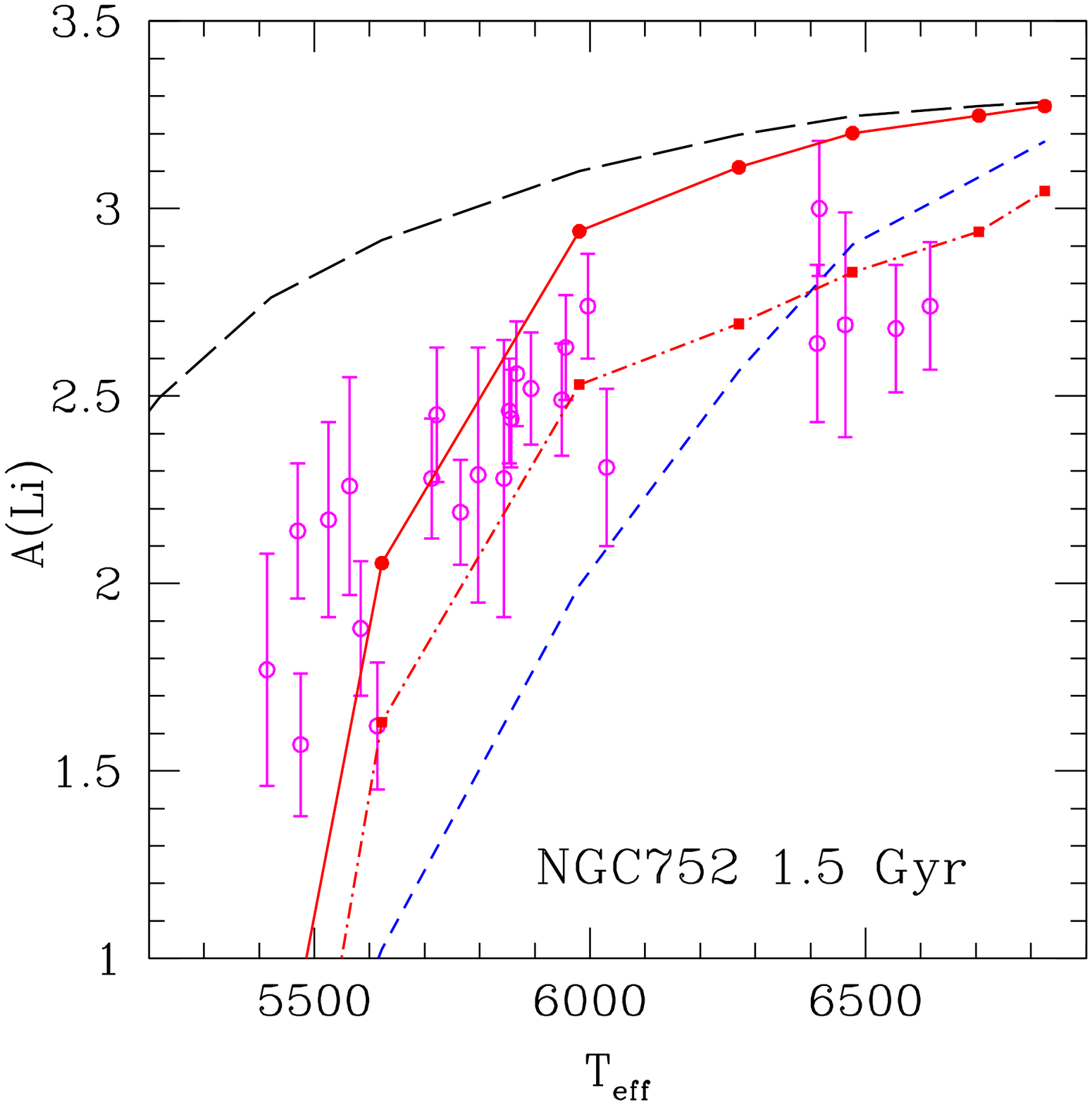}{0.35\textwidth}{\vspace{-1.5cm} (d)}
          \fig{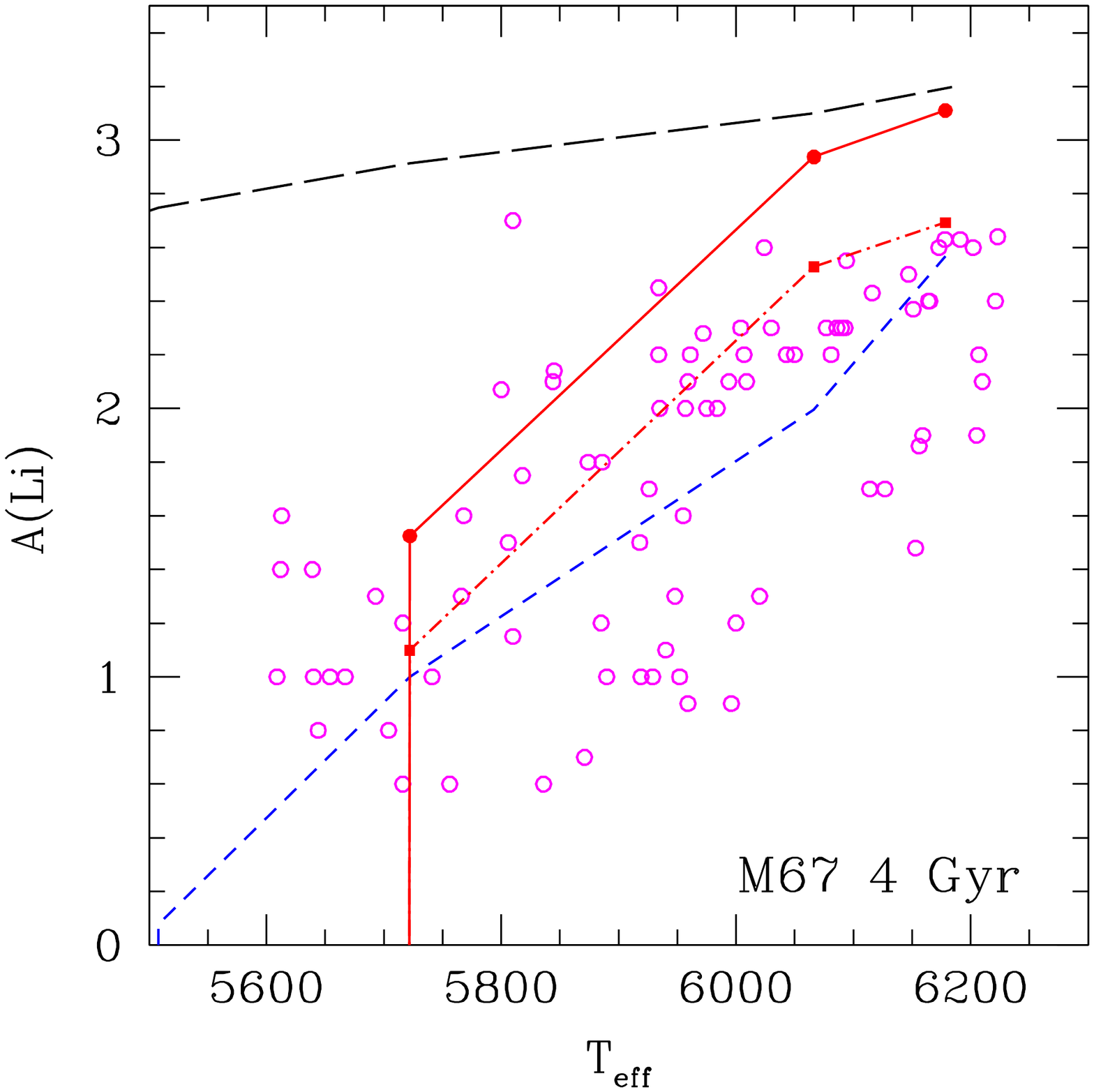}{0.35\textwidth}{\vspace{-1.5cm} (e)}}
          \vspace{-1.5cm}
\caption{Lithium abundance versus effective temperature in different clusters. The observations are for (a) IC2391 (blue diamonds) and IC2602 (magenta dots) from \citet{Randich01}; (b) Pleiades from 
\citet{Barrado16} (magenta diamond) and \citet{Gondoin14} (black dots); (c) Hyades from \citet{Castro16} (magenta circle) and \citet{Cummings17} (blue squares); (d) NGC752 from \citet{Castro16} (magenta circles); (e) M67 from \citet{Castro16} (magenta circles). The different curves correspond to stellar models with masses between 0.85-1.5 $\msun$ with different treatments of overshooting, with the same linestyles and colours as in Figure \ref{fig_tli}. The symbols on the red curves indicate the masses 0.85, 0.9, 0.95, 1, 1.1, 1.2, 1.3, 1.4 and 1.5 $\msun$ (from left to right; note that in (d) and (e) only the symbols for the highest masses are shown, Li being fully depleted in the lowest masses).
\label{fig_teli}}
\end{figure*}

\section{Lithium depletion as a function of time: the role of rotation} 
\label{sect4}

To compare the depletion of lithium as a function of time with observations, we compute a range of stellar models using  $\dpratt$ with limiting width $\dov = 0.35 \Hp$, for stars with masses between 0.85 $\msol$ and 1.5 $\msol$.   
We find that this treatment is unable to explain  Li abundance if applied consistently throughout early and late stages of stellar evolution. 
This is illustrated in Figure \ref{fig_teli}. A comparison between observations and models with mixing at the convective boundary according to  $\dpratt$ shows a significant overestimate of Li depletion for ages  $\simlt$ 2 Gyr.  For older ages,  a conclusion is difficult to reach given the large spread of the data for M67.
We conclude that either the form of the diffusion coefficient, or the limiting width must be allowed to change as the star ages.

To probe this discrepancy, we explore a simple scenario that relies on the effect of rotation on the convective plumes. Recent numerical studies \citep{Ziegler03,Brummell07,Brun17} that focus on the combined effects of convection and rotation suggest that rotation may reduce the penetration depth of convective motions at the convective boundary when the rotation rate is high. Fast rotation may result in tilting of the down-flows and an enhancement of horizontal mixing at the expense of vertical mixing, in the overshoot region. A similar effect can be found in buoyant plumes in deep oceans \citep{Fabregat16}. 
However, there has been so far no quantitative estimate of the effect of rotation on the width of the overshooting layer and on the mixing efficiency in this layer. Inspired by our description of convective boundary mixing and the works above-mentioned, we  suggest a scenario in which mixing at the convective boundary takes place according to our statistical diffusion coefficient  $\dpratt$ defined in \S \ref{sect2}. But compared to the treatment adopted in \S \ref{sect3},  we assume a maximum penetration depth that depends on the rotation rate, with fast rotation strongly limiting the vertical penetration of {the most vigorous convective plumes}. Our assumptions are the following:
\begin{packed_item}
\item{} (i) 
 Below a critical rotation rate $\Omega_{\rm crit}$ (for slow rotators),  mixing takes place down to $\dov \sim 1 \, \Hp$. 
\item{} (ii) Above $\Omega_{\rm crit}$ (for fast rotators), mixing is limited to 0.1 $\Hp$. 
\item{} (iii) We adopt  $\Omega_{\rm crit} = 5  \, \Omega_\odot $, corresponding to a period $P_{\rm crit}=5$ days ($\Omega_\odot=2.9 \cdot 10^{-6}~ \mathrm{rad \cdot s}^{-1}$ or $P_\odot = 25$ days). 
\end{packed_item}

\begin{figure}[h]
\includegraphics[height=14cm]{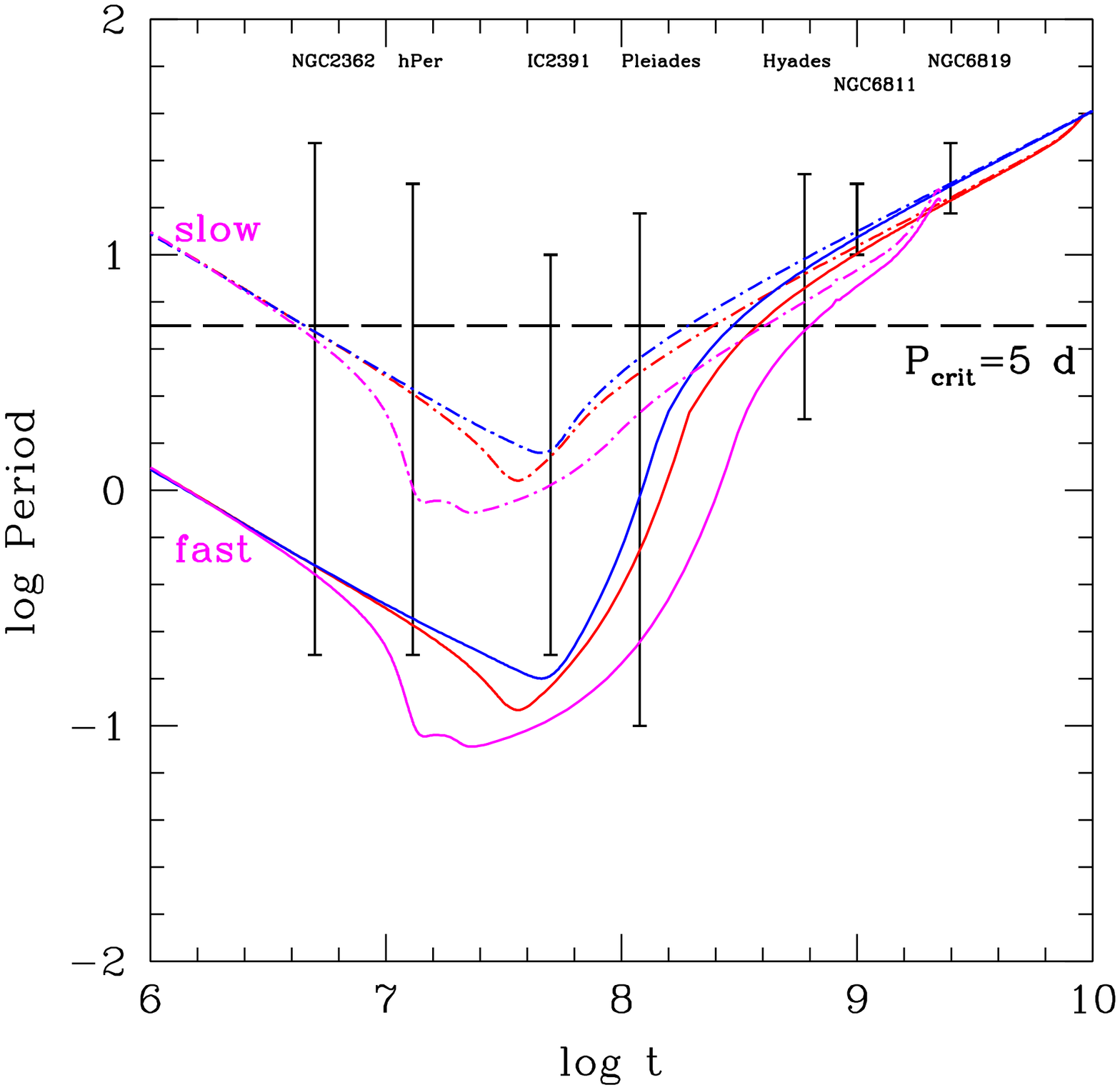}
\caption{Rotation period (in days) versus time (in yr) for initially slow ($P_0$= 12.5 d; dash-dotted lines) and fast  ($P_0$= 1.25 d; solid lines) rotating stars of mass 0.85 $\msol$ (blue), 1 $\msol$ (red) and 1.5 $\msol$ (magenta). The black vertical lines give the range of observed periods in clusters of different ages adopted from \cite{Gallet15}.}
\label{fig_tperiod}
\end{figure}

Presently, the values for $\Omega_{\rm crit}$ and $\dov$ are chosen to broadly reproduce the observations. More work is required to find an underlying physical justification.  To test our scenario, we adopt a simple model to follow the evolution of the rotation rate of our stellar models, assuming solid-body rotation and magnetic braking using Kawaler's law \citep{Kawaler88}  \citep[see details in][]{Bouvier97, Viallet12}.  We assume two different initial rotation periods  to broadly cover the range of
 observed periods in young clusters: slow initial  rotators with a rotation period $P_0$= 12.5 days ($\Omega_0 =  2  \, \Omega_\odot$)
and fast  initial rotators with $P_0$= 1.25 days ($\Omega_0 = 20 \,  \Omega_\odot$).
As the model used provides a reasonable estimate for the evolution of the rotation rate of stars in the considered mass range, and given that the rotation does not directly impact the evolution itself, the details of the rotation model are not critical for the present purpose. 
This is illustrated in Figure \ref{fig_tperiod}, which compares the predicted rotation periods for models between 0.85-1.5 $\msol$ and the range of measured period in clusters of different ages. 

The results for Li depletion, shown in Figure \ref{fig_tli} for a 1 $\msol$ star, are expanded for a range of stellar masses in Figure \ref{fig_teli}. In Figure \ref{fig_tli}, the observed solar Li abundance can be matched with this scenario, assuming a slow initial rotation period. 
The range of masses explored in Figure \ref{fig_teli} shows that the models recover the correct trend of Li depletion as a function of age. The inhibiting effect of rotation on the  vertical penetration of the plume while the star is young and rapidly rotating prevents lithium destruction at early ages, which was previously found with models that do not account for rotation (as illustrated by comparing the red and blue curves in Figure  \ref{fig_tli} ).
The other interesting result is that the models can reproduce the observed correlation between rotation and Li abundance, with the faster rotators being less Li depleted than their slower counterparts (Figure \ref{fig_teli}).

\section{Discussion and Conclusion} 
\label{conclusion}

While models based on the present scenario convincingly recover the observed trend of lithium depletion as a function of age, a close comparison with observations shown in Figure \ref{fig_teli} reveals certain discrepancies. The well-known lithium dip or gap at $\te \sim 6500$K, which is particularly visible for the Hyades, is not reproduced, and is not explained by our scenario \citep[see][and references therein]{Barrado16}.
However, for a simple model that uses a physically-based diffusion coefficient and a simple two-parameter limiting process based on a critical rotation rate $\Omega_{\rm crit}$, and limiting width $\dov$, this recovers the trend in observations remarkably well. Although parameters could be better tuned to match the observations, that is not our objective.  Our goal is to explore a new scenario that could provide an explanation for Li depletion in low-mass and solar-like stars as a function of age, based on a reduced number of assumptions and relying on a physical and statistical model of convective boundary mixing derived from numerical simulations. Even in the simple form presented here, this scenario can explain the observed trend between rotation and depletion,
based on one major assumption, namely that rotation affects the vertical penetration of plumes and thus the mixing efficiency at the convective boundary. We suggest the existence of a threshold $\Omega_{\rm crit}$ above which rotation strongly prevents the vertical penetration of plumes, even the most vigorous ones, and below which rotation has small effects on the most vigorous plumes. Interestingly enough,  the idea of a critical rotation can be found in the theoretical work of \citet{Rudiger01}, which highlights the complex interplay between rotation, turbulence and diffusion and shows how rotation can suppress and deform turbulence, with a net effect of enhancing or reducing chemical mixing depending on the rotation rate.
The strength of our scenario is that its assumptions are testable with hydrodynamic numerical simulations. This is our future goal. We plan to perform the same MUSIC simulations as \citet{Pratt17} for a range of stellar masses and stages of evolution and also include the effects of different rotation rates.  We also plan to explore other observational signatures of our statistical model for mixing at convective boundaries, namely the depletion of beryllium and the heat transport that may affect the temperature profile in the overshooting region, which can be directly compared  with helioseismology and asteroseismology data. 

This work represents a successful effort to directly link multi-dimensional hydrodynamic simulations to stellar evolution models and observations. It demonstrates the potential of our statistical approach to quantitatively derive diffusion coefficients for convective boundary mixing and provides a promising avenue  to improve one-dimensional stellar evolution models.

\acknowledgments
The authors thank Philippe Gondoin for providing his data. This work is funded by the ERC grant No. 320478-TOFU.
It used the DiRAC Complexity system which is part of the STFC DiRAC HPC Facility (funded by BIS National E-Infrastructure capital grant ST/K000373/1 and  Operations grant ST/K0003259/1).
This work also used the University of Exeter Supercomputer ISCA.


\end{document}